\documentclass{article}

\usepackage[dblblindworkshop, final]{neurips_ACA_2025}

\workshoptitle{\\ Workshop on Algorithmic Collective Action (ACA@NeurIPS 2025)}

\usepackage[utf8]{inputenc} 
\usepackage[T1]{fontenc}    
\usepackage{hyperref}       
\usepackage{url}            
\usepackage{booktabs}       
\usepackage{amsfonts}       
\usepackage{nicefrac}       
\usepackage{microtype}      
\usepackage{xcolor}         

\usepackage{enumitem}
\usepackage[most]{tcolorbox} 

\usepackage{microtype}
\usepackage{hyperref}
\usepackage{url}
\usepackage{booktabs}
\usepackage{amsmath}
\usepackage{lineno}

\usepackage{mathtools}
\usepackage{bm}

\title{The Last Vote: A Multi-Stakeholder Framework for Language Model Governance}

\author{%
  Subramanyam Sahoo\thanks{Core Contributor} \\
   Berkeley AI Safety Initiative (BASIS)\\
 UC Berkeley\\
  \texttt{sahoo2vec@gmail.com} \\
 \And
 Aditi Chhawacharia\\
Cyber Physical Systems Lab \\
University of Texas, Dallas \\
\texttt{chhawacharia.aditi@gmail.com} \\
}

\begin{document}

\maketitle

\begin{abstract}
As artificial intelligence systems become increasingly powerful and pervasive, democratic societies face unprecedented challenges in governing these technologies while preserving core democratic values and institutions. This paper presents a comprehensive framework to address the full spectrum of risks that AI poses to democratic societies. Our approach integrates multi-stakeholder participation, civil society engagement, and existing international governance frameworks while introducing novel mechanisms for risk assessment and institutional adaptation. We propose: (1) a seven-category democratic risk taxonomy extending beyond individual-level harms to capture systemic threats, (2) a stakeholder-adaptive Incident Severity Score (ISS) that incorporates diverse perspectives and context-dependent risk factors, and (3) a phased implementation strategy that acknowledges the complex institutional changes required for effective AI governance.
\end{abstract}

Keywords: AI Governance, Democratic Risks, ISS Score

\section{Introduction}

Language model governance remains marked by technocratic reductionism, privileging compliance-oriented risk taxonomies and existential mitigation logics while occluding the constitutive politicality of AI as a socio-technical infrastructure that redistributes epistemic authority and encodes normative commitments \cite{araujo2024understanding}. This depoliticized framing produces legitimacy deficits across both state-centric regulatory instruments and industry self-regulation, which operationalize governance as a problem of optimization rather than democratic authorization. Contemporary interventions, exemplified by the \cite{AIAct2024}'s AI Act and recent U.S. executive orders, largely delimit governance to individualized harms and discrete technical risk vectors, thereby neglecting structural modalities through which scaled generative systems destabilize democratic legitimacy \cite{farnadi2024positioncrackingcodecascading}. Such systems, by virtue of their cascading and path-dependent societal effects, engender forms of procedural erosion that elude capture by extant risk assessment methodologies, underscoring the need for governance architectures attuned to systemic threats beyond the technical-compliance paradigm \cite{feng2023pretraining,diakopoulos2024anticipating}.

\textbf{Democratic AI Governance Framework:} We formalize AI governance as an optimization problem where democratic integrity is the primary objective, not a side-constraint \cite{cooper2024machineunlearningdoesntthink}. This formulation remedies three deficits in existing paradigms: (i) Risk Assessment: shifting from local fairness metrics to systemic analyses of AI–democracy interactions; (ii) Legitimacy: replacing technocratic exclusivity with participatory, binding authority for civil society and affected stakeholders; (iii) Implementation: supplying concrete institutional and procedural mechanisms that instantiate democratic principles as operational rules rather than aspirational norms \cite{bovens2007analysing, arnstein1969ladder} (For more -- Follow Section 3).

\textbf{Key Contributions:} We extend AI governance literature by (i) introducing a seven-category taxonomy of democratic risks spanning individual exclusion to systemic fragility, (ii) formalizing a stakeholder-adaptive Incident Severity Score (ISS) that aggregates heterogeneous utilities into mathematically rigorous governance signals, (iii) proposing a four-phase, six-year implementation roadmap transitioning from voluntary coordination to binding democratic oversight, and (iv) operationalizing deliberative democratic theory through institutionalized co-governance, citizen panels, and sovereignty zones.

\textbf{Paper Structure:} Section 2 positions our work relative to existing frameworks. Section 3 presents our democratic risk taxonomy. Section 4 details the multi-stakeholder governance architecture. Section 5 outlines phased implementation. Section 6 covers monitoring and adaptation. The mathematical ISS framework is detailed in the appendix, providing computational tools for operationalizing democratic oversight. Section 7 and 8 gives a clear-cut horizon for Limitations and Future works correspondingly. Additional technical details and responses to methodological critiques appear in Appendices.

\section{Background and Related Work}

\cite{ovadya2025democraticaipossibledemocracy} demonstrates that technologies are not neutral tools but embody political values and redistribute power within society. The work done by \cite{jasanoff2004states} on the co-production of science and social order shows how technological systems and political institutions mutually constitute each other. This perspective reveals why purely technical approaches to AI governance fail to address AI's constitutive political effects on democratic institutions \cite{Hadfield2025}.

\textbf{Current Governance Limitations:}
Existing governance frameworks predominantly address individual-level harms through fairness constraints and safety protocols \cite{jobin2019global_landscape}. The \textbf{EU AI Act} employs risk stratification by application domain but lacks systematic treatment of systemic democratic effects \cite{walters2023complyingeuaiact}. Industry focused partnerships also largely prioritize individual accountability over collective institutional impacts \cite{pasquale2015black_box}. While previous work has addressed group fairness in machine learning through demographic parity and equalized odds metrics, these approaches still operate within individual-level harm frameworks rather than addressing structural democratic threats \cite{fishkin2018democracy}. \textit{Our work extends beyond both individual and group fairness to examine AI's impact on democratic institutions themselves.}

\textbf{Our Positioning:} We position this work at the intersection of AI governance literature on technology's constitutive political effects and computational democracy research emphasizing participatory institutional design. Our framework diverges from prevailing approaches by treating \textbf{democratic integrity as a primary optimization objective} \cite{sahoo2025the} rather than a constraint, contributing a formal risk taxonomy that extends algorithmic governance theory through operationalizable metrics for democratic impact assessment.

\section{A Comprehensive Risk Taxonomy for Democratic Societies }

Building on the domain structure of \cite{mitAIRiskRepo2024,slattery2025airiskrepositorycomprehensive}, we extend existing risk taxonomies to systematically address AI's threats to democratic institutions. Our taxonomy, derived from democratic theory and historical institutional threats, captures both direct process-level risks and indirect institutional interactions, spanning harms from individual exclusion to systemic collapse \cite{bengio2025internationalaisafetyreport}.

\begin{tcolorbox}[breakable,colback=yellow!10!white,colframe=red!60!black]
\textbf{Discrimination \& Democratic Exclusion.} Beyond individual unfair treatment, AI systems can systematically exclude entire communities from democratic participation. This includes algorithmic discrimination in voting access, civic service delivery, and representation in democratic processes, creating structural barriers to political equality. \cite{hewage2023exploring} showed GPT-based resume screening tools systematically exclude candidates based on linguistic patterns associated with minority communities, institutionalizing bias at scale. \cite{hendrycks2023overviewcatastrophicairisks}

\vspace{0.5\baselineskip}
\textbf{Privacy Erosion \& Democratic Surveillance.} AI-assisted surveillance enables unprecedented monitoring of citizen activities, communications, and political associations, potentially chilling free expression and opposition organizing. \cite{10.1145/3712001} This extends privacy concerns into the realm of democratic participation rights. \cite{Agrawal_2022,liang2018constructing} Curated evidence from China's social credit system demonstrates how surveillance can systematically constrain democratic participation by monitoring and scoring citizen behavior, creating chilling effects on dissent and political organization. Corporate surveillance systems in democracies create similar risks through political tracking and behavioral scoring. \cite{FLI2024AISafetyIndex}

\vspace{0.5\baselineskip}
\textbf{Electoral Misinformation \& Discourse Degradation.} Current models enable computational propaganda, hyper-personalized misinformation campaigns, and systematic degradation of civic discourse quality. \cite{aparicio_de_soto2022constructivism} Unlike general misinformation, these threats specifically target electoral processes and democratic deliberation. Bots powered by model weights \cite{nevo2024securing} can be used to conduct targeted surveys, build voter profiles, infer political preferences from conversational data, and deliver personalized propaganda.

\vspace{0.5\baselineskip}
\textbf{Democratic Manipulation \& Malicious Interference.} Sophisticated actors can weaponize model weights for large-scale electoral interference, voter suppression, and systematic manipulation of democratic processes. This extends beyond individual fraud to coordinated attacks on democratic institutions. \cite{ribeiro2020auditing} documents bot networks amplifying divisive political content, synthetic media campaigns targeting specific voter demographics, and automated systems designed to suppress turnout through coordinated disinformation campaigns. \cite{shah2025approach,bullock2025agigovernmentsfreesocieties}

\vspace{0.5\baselineskip}
\textbf{Civic Participation \& Human Agency Loss.} Algorithmic curation of information environments affects civic engagement through echo-chamber reinforcement, filter-bubble creation, and the delegation of civic decision-making to automated systems, reducing meaningful human participation in democracy. \cite{costanza_chock2020design_justice} Language-model–assisted recommendation systems can promote increasingly extreme political content and create radicalization pathways that undermine democratic discourse norms. \cite{ribeiro2020auditing}

\vspace{0.5\baselineskip}
\textbf{Democratic Power Concentration.} The capital and data requirements for advanced AI concentrate power among a few actors, enabling democratic capture through regulatory influence and technological dependency. \cite{sahoo2024boardwalkempiregenerativeai} shows how democratic institutions can become dependent on private entities for critical functions. Foundation-model concentration creates dependencies when governments adopt these systems for public services, potentially delegating consequential democratic decisions to unaccountable private entities. \cite{reuel2024openproblemstechnicalai,fisher2025politicalneutralityaiimpossible}

\vspace{0.5\baselineskip}
\textbf{Systemic Democratic Fragility.} Complex interactions among new-era models \cite{openai2024openaio1card} can produce emergent behaviors that threaten democratic stability through cascade failures, unintended coordination effects, or systems developing goals misaligned with democratic oversight—representing novel risks to institutional stability. \cite{demirer2019herding,hammond2025multiagentrisksadvancedai}
\end{tcolorbox}

\section{A Multi-Stakeholder Governance Architecture}

Current governance frameworks suffer from what \cite{grek2016expert_moves} et al., identifies as the \textbf{expertocracy} problem: the systematic privileging of technical expertise while relegating other epistemic contributions to symbolic consultation . This produces legitimacy deficits because different stakeholder groups possess forms of knowledge that are non-substitutable and cannot be reduced to purely technical metrics \cite{caddle2025buildingvillagemultistakeholderapproach}.

To address this, our framework institutionalizes seven distinct categories of expertise. Technical practitioners provide feasibility assessments and capability boundaries grounded in real-world deployment contexts. Academic researchers contribute interdisciplinary safety analysis and long-term systemic perspectives unconstrained by immediate commercial pressures \cite{ho2023internationalinstitutionsadvancedai}. Democratic representatives ensure electoral legitimacy and constitutional compatibility, embedding governance processes within democratic accountability structures. Civil society organizations offer public-interest advocacy and long-term value-sensitive oversight \cite{VONROSING2025613}. Industry participants contribute knowledge of market dynamics, competitive pressures, and implementation costs that external regulators often lack. Affected communities provide experiential evidence of algorithmic harms that cannot be captured by audits or simulations khan2025randomnessrepresentationunreliabilityevaluating. Finally, international partners supply coordination capacity across jurisdictions and analysis of AI’s transnational effects on democratic institutions .

We propose a graduated model of participatory governance that integrates these knowledge categories through risk-sensitive forms of involvement \cite{bai2022constitutional,parthasarathy2024participatoryapproachesaidevelopment}. In \textit{low-risk} contexts, governance may rely on enhanced consultation and transparency mechanisms such as public comment periods and hearings. \textit{Medium-risk} scenarios require structured deliberation through citizen panels, stakeholder workshops, and anticipatory technology assessment. \textit{High-risk} applications demand binding co-governance, where stakeholder groups exercise formal decision-making authority, including veto rights and access to appeals processes \cite{ganeri2019epistemic} .

\section{A Phased Implementation Strategy}

Comprehensive AI governance cannot be imposed immediately due to: \textbf{(i) weak public salience of systemic risks before crisis events, (ii) insufficient technical capacity in regulatory agencies, (iii) industry resistance absent competitive incentives for compliance, and (iv) democratic legitimacy deficits when governance precedes stakeholder engagement.} Phased implementation addresses these constraints by: building demonstration effects through visible early successes, accumulating technical capacity through learning-by-doing, creating first-mover advantages that flip industry incentives, and generating political coalitions through early stakeholder inclusion \cite{Bengio_2024}. This approach acknowledges that the primary barrier is not technical feasibility but political economy—institutional transformation requires coalition-building, not just framework specification \cite{reuel2025evaluatesaissocialimpacts}.

\subsection{Foundation Building Phase (0-24 months)} -\textit{Establishing Constitutional Democratic AI Governance}

The foundation building phase prioritizes the establishment of robust constitutional frameworks that define clear stakeholder rights, enforcement mechanisms, and accountability structures for effective governance \cite{priyanshu2024aigovernanceaccountabilityanalysis}. This initial phase focuses on legitimacy building through controlled pilot deployments that test core governance hypotheses in low-risk, high-visibility settings \cite{chaffer2025decentralizedgovernanceautonomousai}. Specifically, municipal bodies must serve as testing grounds for political chatbots and content moderation language models, allowing for real world validation of democratic oversight mechanisms while minimizing systemic risks \cite{Huang_2024}. These pilot programs will generate empirical evidence on stakeholder engagement effectiveness, regulatory compliance costs, and democratic participation outcomes that will inform subsequent phases \cite{Allen2025Roadmap}. The phase concludes with the codification of constitutional principles including due process rights for affected communities, transparency requirements for algorithmic decision making, and appeals mechanisms for automated determinations that impact democratic participation \cite{ribeiro2025effectiveaigovernancereview}.

\subsection{System Integration Phase (24-48 months)}

-\textit{Transitioning to Mandatory Compliance and Risk-Based Oversight}

The system integration phase marks the critical transition from voluntary industry cooperation to mandatory regulatory compliance, with particular emphasis on high-risk applications that directly impact democratic processes \cite{hadfield2023regulatorymarketsfutureai}. All deployments involving \textbf{automated political advertising, synthetic news generation, or voter-targeted conversational agents} will be subject to mandatory "\textbf{\textit{Incident Severity Score (ISS)}}" (Refer Appendix A) assessments conducted by fully operational model safety committees with diverse stakeholder representation. These committees will possess enforcement authority, including the power to require design modifications, impose operational restrictions, or mandate system shutdowns for applications that exceed established risk thresholds \cite{zeng2024airiskcategorizationdecoded}. This phase includes the development of standardized assessment protocols, the training of qualified evaluators, and the establishment of inter-agency coordination mechanisms to ensure consistent application of governance standards across jurisdictions \cite{pazzaglia2025passingturingtestpolitical}. By the conclusion of this phase, the regulatory framework will demonstrate measurable effectiveness in identifying and mitigating high-risk deployments while maintaining democratic legitimacy through transparent, participatory oversight processes.

\subsection{Comprehensive Coverage Phase (48-72 months)}

-\textit{Expanding Regulatory Scope Through Decentralized Democratic Oversight}

This phase extends mandatory governance requirements to medium risk scenarios while adopting the principle of subsidiarity through community based oversight mechanisms. Local community oversight boards, composed of affected stakeholders and technical experts, will assume primary responsibility for evaluating systems with localized impacts, such as educational content generation tools, and community specific content moderation systems \cite{terminassian2025democratizingaigovernancebalancing}. These decentralized bodies will operate within standardized frameworks established during previous phases while retaining authority to adapt governance approaches to local democratic values and community needs \cite{ovadya2025democraticaipossibledemocracy}. The phase emphasizes capacity building through \textbf{comprehensive training programs for community oversight members, the development of accessible technical assessment tools, and the establishment of resource-sharing networks between communities} \cite{10.1093/polsoc/puae022}. This approach ensures that governance parameters scale democratically rather than bureaucratically, maintaining citizen engagement and local accountability as regulatory coverage expands across the ecosystem \cite{reuel2024generativeaineedsadaptive}.

\subsection{Adaptive Governance Phase (72+ months)}

-\textit{Institutionalizing Continuous Democratic Learning and Innovation}

This phase institutionalizes mechanisms for continuous democratic learning and \textbf{governance innovation}, ensuring that regulatory frameworks evolve alongside technological developments and dynamic societal values \cite{kulothungan2025adaptiveaigovernancecomparative}. Governance innovation laboratories will serve as controlled environments for testing novel oversight approaches, stakeholder engagement mechanisms, and risk assessment methodologies before their incorporation into mainstream regulatory practice. These laboratories will operate through partnerships between regulatory agencies, academic institutions, and civil society organizations, generating empirical evidence on governance effectiveness and democratic legitimacy \cite{zhong2025globalaigovernancechallenge}. The phase includes the establishment of systematic processes for updating risk thresholds based on emerging evidence, regular review cycles for stakeholder representation mechanisms, and adaptive procedures for incorporating lessons learned from governance failures or unexpected outcomes \cite{ahern2025newanticipatorygovernanceculture}. This institutionalized learning approach ensures that democratic AI governance remains responsive to technological change while preserving core democratic values and maintaining public trust in regulatory institutions.

Coalition resilience underpins institutional transformation through strategic stakeholder alignment that preempts governance capture while ensuring political sustainability \cite{stańczak2025societalalignmentframeworksimprove}. This approach mobilizes civil society organizations as advocacy coalitions, deploys public education campaigns to build democratic legitimacy, and develops industry partnerships by framing robust governance as market-stabilizing infrastructure. The framework addresses the ``\textbf{democratic deficit problem}'' \cite{azman2011problem} in AI governance by creating self-reinforcing political incentives: early adopters gain competitive advantages through public trust premiums, while compliance costs decrease through economies of scale as participation expands, generating positive feedback loops that sustain democratic governance against technocratic reversion \cite{longpre2025inhouseevaluationenoughrobust}.

\section{Monitoring, Evaluation, and Adaptation}

Governance should be seen as a cybernetic homeostatic system through a tripartite impact monitoring protocol that continuously tracks: (1) democratic health indicators including electoral integrity metrics and civic discourse quality measures, (2) longitudinal social-economic impact assessments quantifying equity shifts and disparate model deployment effects on marginalized populations, and (3) governance system performance metrics evaluating decision-making efficiency and procedural justice (\cite{zwitter2024cybernetic_governance,zaidan2024ai_governance}). This multi-modal stream generates high-fidelity real-time diagnostics that feed directly into dual-architecture adaptation mechanisms: systematic annual threshold recalibration based on empirical outcomes, and dedicated governance innovation labs serving as institutional sandboxes for novel oversight methodologies \cite{salaudeen2025measurementmeaningvaliditycenteredframework}. Accountability is enforced through mandated transparency protocols including public facing dashboards and annual governance reports, while independent statutory oversight bodies conduct external audits to prevent regulatory capture and ensure \textbf{democratic alignment} \cite{hendrycks2025superintelligencestrategyexpertversion}. This architecture transforms static regulatory frameworks into evidence-based adaptive systems capable of responding to emergent socio-technical pathologies while maintaining procedural legitimacy \cite{greenblatt2024alignmentfakinglargelanguage,summerfield2024advancedaisystemsimpact}.

\section{Current Limitations}

The stakeholder weight aggregation mechanism assumes \textit{rational behavior} and may overlook power dynamics or strategic manipulation that characterize actual democratic processes. Additionally, the framework confronts substantial methodological constraints including cultural specificity to Western democratic contexts that limits global applicability and resource-intensive deliberative processes that may exceed organizational capacity. Implementation challenges include assumptions of institutional willingness to adopt multi-stakeholder governance without addressing entrenched interests that benefit from existing technocratic approaches, and limited enforcement mechanisms for compelling compliance from powerful AI companies or state actors who may resist democratic oversight. Technical gaps encompass unclear methodologies for identifying and legitimizing community representatives, raising concerns about the democratic legitimacy of the stakeholders themselves, and potential inability to capture qualitative aspects of democratic harm such as the erosion of civic trust or degradation of deliberative norms that resist quantification.

\section{Future Directions}

Future work must empirically validate the ISS on documented governance failures, track democratic health longitudinally, and adapt the framework cross-culturally to separate universal from local principles. \textbf{Methodological advances should build dynamic online learning systems with causal inference and uncertainty quantification.} Research should integrate governance innovations (hybrid human–AI oversight, decentralized transparency, international coordination), technical extensions (real-time monitoring, adversarial robustness, multi-modal risk assessment), and practical pathways (municipal testbeds, industry incentives, legal integration). Theoretical development drawing on deliberative democracy, critical power analysis, and complexity science is essential to evaluate effectiveness, legitimacy, and scalability, ultimately testing whether multi-stakeholder AI governance can enhance democratic resilience.

\section{Conclusion}
Language model governance is fundamentally a political challenge requiring democratic solutions, not just technical ones. This framework strengthens democracy by creating new mechanisms for participation, accountability, and transparency while providing concrete tools for implementation.
Our ISS metric offers a mathematically rigorous approach to risk assessment that incorporates stakeholder expertise without sacrificing technical precision. The phased implementation strategy provides realistic pathways for institutional transformation while maintaining democratic legitimacy throughout the transition.
The choices made now about language model governance will determine whether these systems strengthen democratic discourse through improved information access or undermine it through manipulation, misinformation, and exclusion. This framework provides tools for ensuring AI serves democratic values rather than subverting them.

\section{Social Impacts Statement}

This work aims to strengthen democratic governance of language models, with several important broader impacts to consider. The proposed governance framework could significantly reshape how societies balance technological innovation with democratic accountability, potentially setting global precedents for managing powerful systems.

\section*{Acknowledgements}

Subramanyam Sahoo would like to thank Stephen Casper (MIT Algorithmic Alignment Group), Jeffrey Andrade (Director, Harvard AISST),  
Chanden Climaco (Harvard), Vinaya Sivakumar (UC Berkeley), Kayla Y Jew (UC Berkeley). He further extends our gratitude to Coby Joseph, Dr.~Aishwarya Saxena, Cristina Schmidt Ibáñez, and Kayla-Leigh Coetzee of the Vista Institute for AI Policy for their valuable discussion during fellowship. He would also like to thank the Apart Lab members — Philip Quirke, Amir Abdullah, and Jacob Haimes — for their support throughout this work.

Both authors would also like to thank the reviewers of the \textit{NeurIPS 2025 Workshop on Algorithmic Collective Action} for their thoughtful feedback and constructive suggestions.

\section*{Spiritual Dedication}

I, Subramanyam Sahoo, dedicate this work to \textbf{Ilya Sutskever}, whose vision and commitment to advancing safe and beneficial AI have profoundly shaped my own path. His ideas inspired me to pursue AI safety, and this work is a small reflection of that influence.

\bibliography{reference}

@misc{hendrycks2025superintelligencestrategyexpertversion,
      title={Superintelligence Strategy: Expert Version}, 
      author={Dan Hendrycks and Eric Schmidt and Alexandr Wang},
      year={2025},
      eprint={2503.05628},
      archivePrefix={arXiv},
      primaryClass={cs.CY},
      url={https://arxiv.org/abs/2503.05628}, 
}

@misc{FLI2024AISafetyIndex,
  author       = {{Future of Life Institute}},
  title        = {AI Experts: Major AI Companies Have Significant Safety Gaps},
  year         = {2024},
  month        = {December},
  url          = {https://futureoflife.org/ai-policy/ai-experts-major-ai-companies-have-significant-safety-gaps/},
  note         = {The 2024 AI Safety Index evaluates six leading AI companies across six safety categories, revealing significant gaps in risk management and control strategies.}
}

@article{Bengio_2024,
   title={Managing extreme AI risks amid rapid progress},
   volume={384},
   ISSN={1095-9203},
   url={http://dx.doi.org/10.1126/science.adn0117},
   DOI={10.1126/science.adn0117},
   number={6698},
   journal={Science},
   publisher={American Association for the Advancement of Science (AAAS)},
   author={Bengio, Yoshua and Hinton, Geoffrey and Yao, Andrew and Song, Dawn and Abbeel, Pieter and Darrell, Trevor and Harari, Yuval Noah and Zhang, Ya-Qin and Xue, Lan and Shalev-Shwartz, Shai and Hadfield, Gillian and Clune, Jeff and Maharaj, Tegan and Hutter, Frank and Baydin, Atılım Güneş and McIlraith, Sheila and Gao, Qiqi and Acharya, Ashwin and Krueger, David and Dragan, Anca and Torr, Philip and Russell, Stuart and Kahneman, Daniel and Brauner, Jan and Mindermann, Sören},
   year={2024},
   month=may, pages={842–845} }

@misc{bengio2025internationalaisafetyreport,
      title={International AI Safety Report}, 
      author={Yoshua Bengio and Sören Mindermann and Daniel Privitera and Tamay Besiroglu and Rishi Bommasani and Stephen Casper and Yejin Choi and Philip Fox and Ben Garfinkel and Danielle Goldfarb and Hoda Heidari and Anson Ho and Sayash Kapoor and Leila Khalatbari and Shayne Longpre and Sam Manning and Vasilios Mavroudis and Mantas Mazeika and Julian Michael and Jessica Newman and Kwan Yee Ng and Chinasa T. Okolo and Deborah Raji and Girish Sastry and Elizabeth Seger and Theodora Skeadas and Tobin South and Emma Strubell and Florian Tramèr and Lucia Velasco and Nicole Wheeler and Daron Acemoglu and Olubayo Adekanmbi and David Dalrymple and Thomas G. Dietterich and Edward W. Felten and Pascale Fung and Pierre-Olivier Gourinchas and Fredrik Heintz and Geoffrey Hinton and Nick Jennings and Andreas Krause and Susan Leavy and Percy Liang and Teresa Ludermir and Vidushi Marda and Helen Margetts and John McDermid and Jane Munga and Arvind Narayanan and Alondra Nelson and Clara Neppel and Alice Oh and Gopal Ramchurn and Stuart Russell and Marietje Schaake and Bernhard Schölkopf and Dawn Song and Alvaro Soto and Lee Tiedrich and Gaël Varoquaux and Andrew Yao and Ya-Qin Zhang and Fahad Albalawi and Marwan Alserkal and Olubunmi Ajala and Guillaume Avrin and Christian Busch and André Carlos Ponce de Leon Ferreira de Carvalho and Bronwyn Fox and Amandeep Singh Gill and Ahmet Halit Hatip and Juha Heikkilä and Gill Jolly and Ziv Katzir and Hiroaki Kitano and Antonio Krüger and Chris Johnson and Saif M. Khan and Kyoung Mu Lee and Dominic Vincent Ligot and Oleksii Molchanovskyi and Andrea Monti and Nusu Mwamanzi and Mona Nemer and Nuria Oliver and José Ramón López Portillo and Balaraman Ravindran and Raquel Pezoa Rivera and Hammam Riza and Crystal Rugege and Ciarán Seoighe and Jerry Sheehan and Haroon Sheikh and Denise Wong and Yi Zeng},
      year={2025},
      eprint={2501.17805},
      archivePrefix={arXiv},
      primaryClass={cs.CY},
      url={https://arxiv.org/abs/2501.17805}, 
}

@misc{hendrycks2023overviewcatastrophicairisks,
      title={An Overview of Catastrophic AI Risks}, 
      author={Dan Hendrycks and Mantas Mazeika and Thomas Woodside},
      year={2023},
      eprint={2306.12001},
      archivePrefix={arXiv},
      primaryClass={cs.CY},
      url={https://arxiv.org/abs/2306.12001}, 
}

@techreport{nevo2024securing,
  title        = {Securing AI Model Weights: Preventing Theft and Misuse of Frontier Models},
  author       = {Sella Nevo and Dan Lahav and Ajay Karpur and Yogev Bar-On and Henry Alexander Bradley and Jeff Alstott},
  institution  = {RAND Corporation},
  number       = {RR-A2849-1},
  year         = {2024},
  month        = {May},
  url          = {https://www.rand.org/pubs/research_reports/RRA2849-1.html},
  note         = {Research Report}
}

@misc{ho2023internationalinstitutionsadvancedai,
      title={International Institutions for Advanced AI}, 
      author={Lewis Ho and Joslyn Barnhart and Robert Trager and Yoshua Bengio and Miles Brundage and Allison Carnegie and Rumman Chowdhury and Allan Dafoe and Gillian Hadfield and Margaret Levi and Duncan Snidal},
      year={2023},
      eprint={2307.04699},
      archivePrefix={arXiv},
      primaryClass={cs.CY},
      url={https://arxiv.org/abs/2307.04699}, 
}

@misc{cooper2024machineunlearningdoesntthink,
      title={Machine Unlearning Doesn't Do What You Think: Lessons for Generative AI Policy, Research, and Practice}, 
      author={A. Feder Cooper and Christopher A. Choquette-Choo and Miranda Bogen and Matthew Jagielski and Katja Filippova and Ken Ziyu Liu and Alexandra Chouldechova and Jamie Hayes and Yangsibo Huang and Niloofar Mireshghallah and Ilia Shumailov and Eleni Triantafillou and Peter Kairouz and Nicole Mitchell and Percy Liang and Daniel E. Ho and Yejin Choi and Sanmi Koyejo and Fernando Delgado and James Grimmelmann and Vitaly Shmatikov and Christopher De Sa and Solon Barocas and Amy Cyphert and Mark Lemley and danah boyd and Jennifer Wortman Vaughan and Miles Brundage and David Bau and Seth Neel and Abigail Z. Jacobs and Andreas Terzis and Hanna Wallach and Nicolas Papernot and Katherine Lee},
      year={2024},
      eprint={2412.06966},
      archivePrefix={arXiv},
      primaryClass={cs.LG},
      url={https://arxiv.org/abs/2412.06966}, 
}

@misc{fisher2025politicalneutralityaiimpossible,
      title={Political Neutrality in AI is Impossible- But Here is How to Approximate it}, 
      author={Jillian Fisher and Ruth E. Appel and Chan Young Park and Yujin Potter and Liwei Jiang and Taylor Sorensen and Shangbin Feng and Yulia Tsvetkov and Margaret E. Roberts and Jennifer Pan and Dawn Song and Yejin Choi},
      year={2025},
      eprint={2503.05728},
      archivePrefix={arXiv},
      primaryClass={cs.CY},
      url={https://arxiv.org/abs/2503.05728}, 
}

@misc{reuel2024openproblemstechnicalai,
      title={Open Problems in Technical AI Governance}, 
      author={Anka Reuel and Ben Bucknall and Stephen Casper and Tim Fist and Lisa Soder and Onni Aarne and Lewis Hammond and Lujain Ibrahim and Alan Chan and Peter Wills and Markus Anderljung and Ben Garfinkel and Lennart Heim and Andrew Trask and Gabriel Mukobi and Rylan Schaeffer and Mauricio Baker and Sara Hooker and Irene Solaiman and Alexandra Sasha Luccioni and Nitarshan Rajkumar and Nicolas Moës and Jeffrey Ladish and Neel Guha and Jessica Newman and Yoshua Bengio and Tobin South and Alex Pentland and Sanmi Koyejo and Mykel J. Kochenderfer and Robert Trager},
      year={2024},
      eprint={2407.14981},
      archivePrefix={arXiv},
      primaryClass={cs.CY},
      url={https://arxiv.org/abs/2407.14981}, 
}

@misc{shah2025approach,
    title={An Approach to Technical AGI Safety and Security},
    author={Rohin Shah and Alex Irpan and Alexander Matt Turner and Anna Wang and Arthur Conmy and David Lindner and Jonah Brown-Cohen and Lewis Ho and Neel Nanda and Raluca Ada Popa and Rishub Jain and Rory Greig and Samuel Albanie and Scott Emmons and Sebastian Farquhar and Sébastien Krier and Senthooran Rajamanoharan and Sophie Bridgers and Tobi Ijitoye and Tom Everitt and Victoria Krakovna and Vikrant Varma and Vladimir Mikulik and Zachary Kenton and Dave Orr and Shane Legg and Noah Goodman and Allan Dafoe and Four Flynn and Anca Dragan},
    year={2025},
    eprint={2504.01849},
    archivePrefix={arXiv},
    primaryClass={cs.AI}
}

@misc{bullock2025agigovernmentsfreesocieties,
      title={AGI, Governments, and Free Societies}, 
      author={Justin B. Bullock and Samuel Hammond and Seb Krier},
      year={2025},
      eprint={2503.05710},
      archivePrefix={arXiv},
      primaryClass={cs.CY},
      url={https://arxiv.org/abs/2503.05710}, 
}

@misc{longpre2025inhouseevaluationenoughrobust,
      title={In-House Evaluation Is Not Enough: Towards Robust Third-Party Flaw Disclosure for General-Purpose AI}, 
      author={Shayne Longpre and Kevin Klyman and Ruth E. Appel and Sayash Kapoor and Rishi Bommasani and Michelle Sahar and Sean McGregor and Avijit Ghosh and Borhane Blili-Hamelin and Nathan Butters and Alondra Nelson and Amit Elazari and Andrew Sellars and Casey John Ellis and Dane Sherrets and Dawn Song and Harley Geiger and Ilona Cohen and Lauren McIlvenny and Madhulika Srikumar and Mark M. Jaycox and Markus Anderljung and Nadine Farid Johnson and Nicholas Carlini and Nicolas Miailhe and Nik Marda and Peter Henderson and Rebecca S. Portnoff and Rebecca Weiss and Victoria Westerhoff and Yacine Jernite and Rumman Chowdhury and Percy Liang and Arvind Narayanan},
      year={2025},
      eprint={2503.16861},
      archivePrefix={arXiv},
      primaryClass={cs.AI},
      url={https://arxiv.org/abs/2503.16861}, 
}

@misc{greenblatt2024alignmentfakinglargelanguage,
      title={Alignment faking in large language models}, 
      author={Ryan Greenblatt and Carson Denison and Benjamin Wright and Fabien Roger and Monte MacDiarmid and Sam Marks and Johannes Treutlein and Tim Belonax and Jack Chen and David Duvenaud and Akbir Khan and Julian Michael and Sören Mindermann and Ethan Perez and Linda Petrini and Jonathan Uesato and Jared Kaplan and Buck Shlegeris and Samuel R. Bowman and Evan Hubinger},
      year={2024},
      eprint={2412.14093},
      archivePrefix={arXiv},
      primaryClass={cs.AI},
      url={https://arxiv.org/abs/2412.14093}, 
}

@article{AIAct2024,
  title = {Regulation (EU) 2024/1689 of the European Parliament and of the Council of 13 June 2024 laying down harmonised rules on artificial intelligence and amending certain Union legislative acts (Artificial Intelligence Act)},
  author = {{European Union}},
  journal = {Official Journal of the European Union},
  volume = {L 2024/1689},
  year = {2024},
  month = {July},
  day = {12},
  pages = {1--168},
  url = {https://eur-lex.europa.eu/eli/reg/2024/1689/oj/eng},
  note = {Published in the Official Journal on 12 July 2024}
}

@techreport{araujo2024understanding,
  author      = {Renan Araujo and Kristina Fort and Oliver Guest},
  title       = {Understanding the First Wave of AI Safety Institutes: Characteristics, Functions, and Challenges},
  institution = {Institute for AI Policy and Strategy},
  year        = {2024},
  month       = {October},
  url         = {https://www.iaps.ai/research/understanding-aisis},
  note        = {Accessed: 2025-05-08}
}

@book{costanza_chock2020design_justice,
  author       = {Costanza‐Chock, Sasha},
  title        = {Design Justice: Community‑Led Practices to Build the Worlds We Need},
  publisher    = {MIT Press},
  year         = {2020},
  address      = {Cambridge, MA},
  isbn         = {9780262043450},
  note         = {Print and e‑book editions}
}

@article{bovens2007analysing,
  author       = {Bovens, Mark},
  title        = {Analysing and Assessing Accountability: A Conceptual Framework},
  journal      = {European Law Journal},
  volume       = {13},
  number       = {4},
  pages        = {447--468},
  year         = {2007},
  doi          = {10.1111/j.1468-0386.2007.00378.x},
  note         = {First published online 7 June 2007}
}

@article{arnstein1969ladder,
  author       = {Arnstein, Sherry R.},
  title        = {A Ladder of Citizen Participation},
  journal      = {Journal of the American Institute of Planners},
  volume       = {35},
  number       = {4},
  pages        = {216--224},
  year         = {1969},
  doi          = {10.1080/01944366908977225},
  note         = {Published July 1969}  
}

@article{demirer2019herding,
  title={Herding and flash events: evidence from the 2010 flash crash},
  author={Demirer, R{\i}za and Leggio, Karyl B and Lien, Donald},
  journal={Finance Research Letters},
  volume={31},
  year={2019},
  publisher={Elsevier}
}

@article{ganeri2019epistemic,
  title={Epistemic pluralism: From systems to stances},
  author={Ganeri, Jonardon},
  journal={Journal of the American Philosophical Association},
  volume={5},
  number={1},
  pages={1--21},
  year={2019},
  publisher={Cambridge University Press}
}

@misc{parthasarathy2024participatoryapproachesaidevelopment,
      title={Participatory Approaches in AI Development and Governance: A Principled Approach}, 
      author={Ambreesh Parthasarathy and Aditya Phalnikar and Ameen Jauhar and Dhruv Somayajula and Gokul S Krishnan and Balaraman Ravindran},
      year={2024},
      eprint={2407.13100},
      archivePrefix={arXiv},
      primaryClass={cs.CY},
      url={https://arxiv.org/abs/2407.13100}, 
}

@article{diakopoulos2024anticipating,
  title={Anticipating and Addressing the Ethical Implications of Deepfakes in the Newsroom},
  author={Diakopoulos, Nicholas and Johnson, David},
  journal={Edward Elgar Publishing (pre‑print)},
  year={2024},
  note={Northwestern Computational Journalism Lab},
  url={https://www.cj-lab.org}
}

@inproceedings{feng2023pretraining,
  title={From Pretraining Data to Language Models to Downstream Tasks: Tracking the Trails of Political Biases Leading to Unfair NLP Models},
  author={Feng, Shangbin and Park, Chan Young and Liu, Yuhan and Tsvetkov, Yulia},
  booktitle={Proceedings of the 61st Annual Meeting of the Association for Computational Linguistics (Volume 1: Long Papers)},
  pages={11737--11762},
  year={2023},
  address={Toronto, Canada},
  publisher={ACL},
  doi={10.18653/v1/2023.acl-long.656},
  url={https://aclanthology.org/2023.acl-long.656}
}

@misc{openai2024openaio1card,
      title={OpenAI o1 System Card}, 
      author={OpenAI and : and Aaron Jaech and Adam Kalai and Adam Lerer and Adam Richardson and Ahmed El-Kishky and Aiden Low and Alec Helyar and Aleksander Madry and Alex Beutel and Alex Carney and Alex Iftimie and Alex Karpenko and Alex Tachard Passos and Alexander Neitz and Alexander Prokofiev and Alexander Wei and Allison Tam and Ally Bennett and Ananya Kumar and Andre Saraiva and Andrea Vallone and Andrew Duberstein and Andrew Kondrich and Andrey Mishchenko and Andy Applebaum and Angela Jiang and Ashvin Nair and Barret Zoph and Behrooz Ghorbani and Ben Rossen and Benjamin Sokolowsky and Boaz Barak and Bob McGrew and Borys Minaiev and Botao Hao and Bowen Baker and Brandon Houghton and Brandon McKinzie and Brydon Eastman and Camillo Lugaresi and Cary Bassin and Cary Hudson and Chak Ming Li and Charles de Bourcy and Chelsea Voss and Chen Shen and Chong Zhang and Chris Koch and Chris Orsinger and Christopher Hesse and Claudia Fischer and Clive Chan and Dan Roberts and Daniel Kappler and Daniel Levy and Daniel Selsam and David Dohan and David Farhi and David Mely and David Robinson and Dimitris Tsipras and Doug Li and Dragos Oprica and Eben Freeman and Eddie Zhang and Edmund Wong and Elizabeth Proehl and Enoch Cheung and Eric Mitchell and Eric Wallace and Erik Ritter and Evan Mays and Fan Wang and Felipe Petroski Such and Filippo Raso and Florencia Leoni and Foivos Tsimpourlas and Francis Song and Fred von Lohmann and Freddie Sulit and Geoff Salmon and Giambattista Parascandolo and Gildas Chabot and Grace Zhao and Greg Brockman and Guillaume Leclerc and Hadi Salman and Haiming Bao and Hao Sheng and Hart Andrin and Hessam Bagherinezhad and Hongyu Ren and Hunter Lightman and Hyung Won Chung and Ian Kivlichan and Ian O'Connell and Ian Osband and Ignasi Clavera Gilaberte and Ilge Akkaya and Ilya Kostrikov and Ilya Sutskever and Irina Kofman and Jakub Pachocki and James Lennon and Jason Wei and Jean Harb and Jerry Twore and Jiacheng Feng and Jiahui Yu and Jiayi Weng and Jie Tang and Jieqi Yu and Joaquin Quiñonero Candela and Joe Palermo and Joel Parish and Johannes Heidecke and John Hallman and John Rizzo and Jonathan Gordon and Jonathan Uesato and Jonathan Ward and Joost Huizinga and Julie Wang and Kai Chen and Kai Xiao and Karan Singhal and Karina Nguyen and Karl Cobbe and Katy Shi and Kayla Wood and Kendra Rimbach and Keren Gu-Lemberg and Kevin Liu and Kevin Lu and Kevin Stone and Kevin Yu and Lama Ahmad and Lauren Yang and Leo Liu and Leon Maksin and Leyton Ho and Liam Fedus and Lilian Weng and Linden Li and Lindsay McCallum and Lindsey Held and Lorenz Kuhn and Lukas Kondraciuk and Lukasz Kaiser and Luke Metz and Madelaine Boyd and Maja Trebacz and Manas Joglekar and Mark Chen and Marko Tintor and Mason Meyer and Matt Jones and Matt Kaufer and Max Schwarzer and Meghan Shah and Mehmet Yatbaz and Melody Y. Guan and Mengyuan Xu and Mengyuan Yan and Mia Glaese and Mianna Chen and Michael Lampe and Michael Malek and Michele Wang and Michelle Fradin and Mike McClay and Mikhail Pavlov and Miles Wang and Mingxuan Wang and Mira Murati and Mo Bavarian and Mostafa Rohaninejad and Nat McAleese and Neil Chowdhury and Neil Chowdhury and Nick Ryder and Nikolas Tezak and Noam Brown and Ofir Nachum and Oleg Boiko and Oleg Murk and Olivia Watkins and Patrick Chao and Paul Ashbourne and Pavel Izmailov and Peter Zhokhov and Rachel Dias and Rahul Arora and Randall Lin and Rapha Gontijo Lopes and Raz Gaon and Reah Miyara and Reimar Leike and Renny Hwang and Rhythm Garg and Robin Brown and Roshan James and Rui Shu and Ryan Cheu and Ryan Greene and Saachi Jain and Sam Altman and Sam Toizer and Sam Toyer and Samuel Miserendino and Sandhini Agarwal and Santiago Hernandez and Sasha Baker and Scott McKinney and Scottie Yan and Shengjia Zhao and Shengli Hu and Shibani Santurkar and Shraman Ray Chaudhuri and Shuyuan Zhang and Siyuan Fu and Spencer Papay and Steph Lin and Suchir Balaji and Suvansh Sanjeev and Szymon Sidor and Tal Broda and Aidan Clark and Tao Wang and Taylor Gordon and Ted Sanders and Tejal Patwardhan and Thibault Sottiaux and Thomas Degry and Thomas Dimson and Tianhao Zheng and Timur Garipov and Tom Stasi and Trapit Bansal and Trevor Creech and Troy Peterson and Tyna Eloundou and Valerie Qi and Vineet Kosaraju and Vinnie Monaco and Vitchyr Pong and Vlad Fomenko and Weiyi Zheng and Wenda Zhou and Wes McCabe and Wojciech Zaremba and Yann Dubois and Yinghai Lu and Yining Chen and Young Cha and Yu Bai and Yuchen He and Yuchen Zhang and Yunyun Wang and Zheng Shao and Zhuohan Li},
      year={2024},
      eprint={2412.16720},
      archivePrefix={arXiv},
      primaryClass={cs.AI},
      url={https://arxiv.org/abs/2412.16720}, 
}

@article{bai2022constitutional,
  title={Constitutional AI: Harmlessness from AI Feedback},
  author={Bai, Yuntao and Bowman, Zac and Hatfield-Dodds, Ben and Mann, Dario and Amodei, Nicholas and McCandlish, Sam and Brown, Tom and Kaplan, Jared and others},
  journal={arXiv preprint arXiv:2212.08073},
  year={2022},
  note={Dec 15 2022},
  url={https://arxiv.org/abs/2212.08073}
}

@article{hewage2023exploring,
  author    = {Hewage, A.},
  title     = {Exploring the Applicability of Artificial Intelligence in Recruitment and Selection Processes: A Focus on the Recruitment Phase},
  journal   = {Journal of Human Resource and Sustainability Studies},
  volume    = {11},
  pages     = {603--634},
  year      = {2023},
  doi       = {10.4236/jhrss.2023.113034}
}

@misc{mitAIRiskRepo2024,
  author = {{MIT AI Risk Repository}},
  title = {Domain Taxonomy of {AI} Risks},
  year = {2024},
  url = {https://airisk.mit.edu},
  note = {Accessed: 2025-01-15}
}

@article{liang2018constructing,
  title={Constructing a data‐driven society: China's social credit system as a state surveillance infrastructure},
  author={Liang, Fan and Das, Vishnupriya and Kostyuk, Nadiya and Hussain, Muzammil M.},
  journal={Policy \& Internet},
  volume={10},
  number={4},
  pages={415--453},
  year={2018},
  publisher={Wiley Online Library},
  doi={10.1002/poi3.183}
}

@article{aparicio_de_soto2022constructivism,
  author    = {Aparicio de Soto, Jesús},
  title     = {The Constructivism of Social Discourse: Toward a Contemporaneous Understanding of Knowledge},
  journal   = {Open Journal of Philosophy},
  volume    = {12},
  number    = {3},
  pages     = {376--396},
  year      = {2022},
  publisher = {Scientific Research Publishing, Inc.},
  doi       = {10.4236/ojpp.2022.123025},
  note      = {Published August 5, 2022}
}

@inproceedings{ribeiro2020auditing,
  author    = {Horta Ribeiro, Manoel and Ottoni, Raphael and West, Robert and Almeida, Virg\'ilio A. F. and Meira Jr., Wagner},
  title     = {Auditing Radicalization Pathways on YouTube},
  booktitle = {Proceedings of the 2020 Conference on Fairness, Accountability, and Transparency (FAT* 2020)},
  pages     = {131--141},
  year      = {2020},
  address   = {Barcelona, Spain},
  publisher = {Association for Computing Machinery},
  doi       = {10.1145/3351095.3372879}
}

@article{jobin2019global_landscape,
  author       = {Jobin, Anna and Ienca, Marcello and Vayena, Effy},
  title        = {The global landscape of AI ethics guidelines},
  journal      = {Nature Machine Intelligence},
  volume       = {1},
  number       = {9},
  pages        = {389--399},
  year         = {2019}
}

@book{jasanoff2004states,
  author       = {Jasanoff, Sheila},
  title        = {States of Knowledge: The Co‑Production of Science and Social Order},
  publisher    = {Routledge},
  year         = {2004},
  address      = {London}
}

@book{fishkin2018democracy,
  author       = {Fishkin, James S.},
  title        = {Democracy When the People Are Thinking},
  publisher    = {Oxford University Press},
  year         = {2018},
  address      = {Oxford}
}

@book{pasquale2015black_box,
  author       = {Pasquale, Frank},
  title        = {The Black Box Society},
  publisher    = {Harvard University Press},
  year         = {2015},
  address      = {Cambridge, MA}
}

@misc{sahoo2024boardwalkempiregenerativeai,
      title={Boardwalk Empire: How Generative AI is Revolutionizing Economic Paradigms}, 
      author={Subramanyam Sahoo and Kamlesh Dutta},
      year={2024},
      eprint={2410.15212},
      archivePrefix={arXiv},
      primaryClass={cs.CE},
      url={https://arxiv.org/abs/2410.15212}, 
}

@article{azman2011problem,
  title={The problem of “democratic deficit” in the European Union},
  author={Azman, K{\"u}bra Dilek},
  journal={International Journal of Humanities and Social Science},
  volume={1},
  number={5},
  pages={242--250},
  year={2011}
}

@incollection{grek2016expert_moves,
  author       = {Grek, Sotiria},
  title        = {Expert moves: International comparative testing and the rise of expertocracy},
  booktitle    = {Testing Regimes, Accountabilities and Education Policy},
  editor       = {Lingard, Bob and Martino, Wayne and Rezai‑Rashti, Goli},
  edition      = {1st},
  pages        = {15},
  publisher    = {Routledge},
  address      = {London},
  year         = {2016},
  isbn         = {9781315666082},
  series       = {},
}

@incollection{VONROSING2025613,
title = {Chapter 40 - Sustainability board with veto rights},
editor = {Mark {von Rosing}},
booktitle = {The Sustainability Handbook, Volume 1},
publisher = {Elsevier},
pages = {613-620},
year = {2025},
isbn = {978-0-323-90110-9},
doi = {https://doi.org/10.1016/B978-0-323-90110-9.00010-6},
url = {https://www.sciencedirect.com/science/article/pii/B9780323901109000106},
author = {Mark {von Rosing} and Lesley Shepperson and Hanka Czichos}
}

@article{zaidan2024ai_governance,
  author       = {Zaidan, Esmat and Ibrahim, Imad Antoine},
  title        = {AI Governance in a Complex and Rapidly Changing Regulatory Landscape: A Global Perspective},
  journal      = {Humanities and Social Sciences Communications},
  volume       = {11},
  number       = {1},
  pages        = {1121},
  year         = {2024},
  doi          = {10.1057/s41599-024-03560-x},
  url          = {https://doi.org/10.1057/s41599-024-03560-x},
  note         = {Received 27 January 2024; Accepted 31 July 2024; Published 1 September 2024}
}

@article{zwitter2024cybernetic_governance,
  author       = {Zwitter, Andrej},
  title        = {Cybernetic governance: implications of technology convergence on governance convergence},
  journal      = {Ethics and Information Technology},
  volume       = {26},
  number       = {2},
  pages        = {1--13},
  year         = {2024},
  doi          = {10.1007/s10676-024-09763-9},
  url          = {https://doi.org/10.1007/s10676-024-09763-9},
  note         = {Published 28 March 2024}
}

@misc{reuel2024generativeaineedsadaptive,
      title={Generative AI Needs Adaptive Governance}, 
      author={Anka Reuel and Trond Arne Undheim},
      year={2024},
      eprint={2406.04554},
      archivePrefix={arXiv},
      primaryClass={cs.CY},
      url={https://arxiv.org/abs/2406.04554}, 
}

@article{10.1093/polsoc/puae022,
    author = {Ulnicane, Inga},
    title = {Governance fix? Power and politics in controversies about governing generative AI},
    journal = {Policy and Society},
    volume = {44},
    number = {1},
    pages = {70-84},
    year = {2024},
    month = {07},    
    issn = {1449-4035},
    doi = {10.1093/polsoc/puae022},
    url = {https://doi.org/10.1093/polsoc/puae022},
    eprint = {https://academic.oup.com/policyandsociety/article-pdf/44/1/70/58386990/puae022.pdf},
}

@article{Allen2025Roadmap,
  author       = {Allen, D. and Hubbard, S. and Lim, W. and et al.},
  title        = {A roadmap for governing AI: technology governance and power-sharing liberalism},
  journal      = {AI Ethics},
  volume       = {5},
  pages        = {3355--3377},
  year         = {2025},
  doi          = {10.1007/s43681-024-00635-y},
  url          = {https://doi.org/10.1007/s43681-024-00635-y},
}

@misc{slattery2025airiskrepositorycomprehensive,
      title={The AI Risk Repository: A Comprehensive Meta-Review, Database, and Taxonomy of Risks From Artificial Intelligence}, 
      author={Peter Slattery and Alexander K. Saeri and Emily A. C. Grundy and Jess Graham and Michael Noetel and Risto Uuk and James Dao and Soroush Pour and Stephen Casper and Neil Thompson},
      year={2025},
      eprint={2408.12622},
      archivePrefix={arXiv},
      primaryClass={cs.AI},
      url={https://arxiv.org/abs/2408.12622}, 
}

@misc{salaudeen2025measurementmeaningvaliditycenteredframework,
      title={Measurement to Meaning: A Validity-Centered Framework for AI Evaluation}, 
      author={Olawale Salaudeen and Anka Reuel and Ahmed Ahmed and Suhana Bedi and Zachary Robertson and Sudharsan Sundar and Ben Domingue and Angelina Wang and Sanmi Koyejo},
      year={2025},
      eprint={2505.10573},
      archivePrefix={arXiv},
      primaryClass={cs.CY},
      url={https://arxiv.org/abs/2505.10573}, 
}

@misc{summerfield2024advancedaisystemsimpact,
      title={How will advanced AI systems impact democracy?}, 
      author={Christopher Summerfield and Lisa Argyle and Michiel Bakker and Teddy Collins and Esin Durmus and Tyna Eloundou and Iason Gabriel and Deep Ganguli and Kobi Hackenburg and Gillian Hadfield and Luke Hewitt and Saffron Huang and Helene Landemore and Nahema Marchal and Aviv Ovadya and Ariel Procaccia and Mathias Risse and Bruce Schneier and Elizabeth Seger and Divya Siddarth and Henrik Skaug Sætra and MH Tessler and Matthew Botvinick},
      year={2024},
      eprint={2409.06729},
      archivePrefix={arXiv},
      primaryClass={cs.CY},
      url={https://arxiv.org/abs/2409.06729}, 
}

@misc{stańczak2025societalalignmentframeworksimprove,
      title={Societal Alignment Frameworks Can Improve LLM Alignment}, 
      author={Karolina Stańczak and Nicholas Meade and Mehar Bhatia and Hattie Zhou and Konstantin Böttinger and Jeremy Barnes and Jason Stanley and Jessica Montgomery and Richard Zemel and Nicolas Papernot and Nicolas Chapados and Denis Therien and Timothy P. Lillicrap and Ana Marasović and Sylvie Delacroix and Gillian K. Hadfield and Siva Reddy},
      year={2025},
      eprint={2503.00069},
      archivePrefix={arXiv},
      primaryClass={cs.CY},
      url={https://arxiv.org/abs/2503.00069}, 
}

@Inbook{Hadfield2025,
author="Hadfield, Gillian K.
and Bernier, Alexander",
title="Revisiting the Many Legal Institutions that Support Contractual Commitments in a Globalized World",
bookTitle="Handbook of New Institutional Economics",
year="2025",
publisher="Springer Nature Switzerland",
address="Cham",
pages="267--292",
isbn="978-3-031-50810-3",
doi="10.1007/978-3-031-50810-3_12",
url="https://doi.org/10.1007/978-3-031-50810-3_12"
}

@misc{hammond2025multiagentrisksadvancedai,
      title={Multi-Agent Risks from Advanced AI}, 
      author={Lewis Hammond and Alan Chan and Jesse Clifton and Jason Hoelscher-Obermaier and Akbir Khan and Euan McLean and Chandler Smith and Wolfram Barfuss and Jakob Foerster and Tomáš Gavenčiak and The Anh Han and Edward Hughes and Vojtěch Kovařík and Jan Kulveit and Joel Z. Leibo and Caspar Oesterheld and Christian Schroeder de Witt and Nisarg Shah and Michael Wellman and Paolo Bova and Theodor Cimpeanu and Carson Ezell and Quentin Feuillade-Montixi and Matija Franklin and Esben Kran and Igor Krawczuk and Max Lamparth and Niklas Lauffer and Alexander Meinke and Sumeet Motwani and Anka Reuel and Vincent Conitzer and Michael Dennis and Iason Gabriel and Adam Gleave and Gillian Hadfield and Nika Haghtalab and Atoosa Kasirzadeh and Sébastien Krier and Kate Larson and Joel Lehman and David C. Parkes and Georgios Piliouras and Iyad Rahwan},
      year={2025},
      eprint={2502.14143},
      archivePrefix={arXiv},
      primaryClass={cs.MA},
      url={https://arxiv.org/abs/2502.14143}, 
}

@article{10.1145/3712001,
author = {Das, Badhan Chandra and Amini, M. Hadi and Wu, Yanzhao},
title = {Security and Privacy Challenges of Large Language Models: A Survey},
year = {2025},
issue_date = {June 2025},
publisher = {Association for Computing Machinery},
address = {New York, NY, USA},
volume = {57},
number = {6},
issn = {0360-0300},
url = {https://doi.org/10.1145/3712001},
doi = {10.1145/3712001},
journal = {ACM Comput. Surv.},
month = feb,
articleno = {152},
numpages = {39}
}

@misc{caddle2025buildingvillagemultistakeholderapproach,
      title={Building a Village: A Multi-stakeholder Approach to Open Innovation and Shared Governance to Promote Youth Online Safety}, 
      author={Xavier V. Caddle and Sarvech Qadir and Charles Hughes and Elizabeth A. Sweigart and Jinkyung Katie Park and Pamela J. Wisniewski},
      year={2025},
      eprint={2504.03971},
      archivePrefix={arXiv},
      primaryClass={cs.CY},
      url={https://arxiv.org/abs/2504.03971}, 
}

@misc{walters2023complyingeuaiact,
      title={Complying with the EU AI Act}, 
      author={Jacintha Walters and Diptish Dey and Debarati Bhaumik and Sophie Horsman},
      year={2023},
      eprint={2307.10458},
      archivePrefix={arXiv},
      primaryClass={cs.AI},
      url={https://arxiv.org/abs/2307.10458}, 
}

@misc{farnadi2024positioncrackingcodecascading,
      title={Position: Cracking the Code of Cascading Disparity Towards Marginalized Communities}, 
      author={Golnoosh Farnadi and Mohammad Havaei and Negar Rostamzadeh},
      year={2024},
      eprint={2406.01757},
      archivePrefix={arXiv},
      primaryClass={cs.LG},
      url={https://arxiv.org/abs/2406.01757}, 
}

@misc{chaffer2025decentralizedgovernanceautonomousai,
      title={Decentralized Governance of Autonomous AI Agents}, 
      author={Tomer Jordi Chaffer and Charles von Goins II and Bayo Okusanya and Dontrail Cotlage and Justin Goldston},
      year={2025},
      eprint={2412.17114},
      archivePrefix={arXiv},
      primaryClass={cs.AI},
      url={https://arxiv.org/abs/2412.17114}, 
}

@inproceedings{Huang_2024, series={FAccT ’24},
   title={Collective Constitutional AI: Aligning a Language Model with Public Input},
   url={http://dx.doi.org/10.1145/3630106.3658979},
   DOI={10.1145/3630106.3658979},
   booktitle={The 2024 ACM Conference on Fairness, Accountability, and Transparency},
   publisher={ACM},
   author={Huang, Saffron and Siddarth, Divya and Lovitt, Liane and Liao, Thomas I. and Durmus, Esin and Tamkin, Alex and Ganguli, Deep},
   year={2024},
   month=jun, pages={1395–1417},
   collection={FAccT ’24} }

@misc{priyanshu2024aigovernanceaccountabilityanalysis,
      title={AI Governance and Accountability: An Analysis of Anthropic's Claude}, 
      author={Aman Priyanshu and Yash Maurya and Zuofei Hong},
      year={2024},
      eprint={2407.01557},
      archivePrefix={arXiv},
      primaryClass={cs.CY},
      url={https://arxiv.org/abs/2407.01557}, 
}

@misc{ribeiro2025effectiveaigovernancereview,
      title={Toward Effective AI Governance: A Review of Principles}, 
      author={Danilo Ribeiro and Thayssa Rocha and Gustavo Pinto and Bruno Cartaxo and Marcelo Amaral and Nicole Davila and Ana Camargo},
      year={2025},
      eprint={2505.23417},
      archivePrefix={arXiv},
      primaryClass={cs.SE},
      url={https://arxiv.org/abs/2505.23417}, 
}

@misc{hadfield2023regulatorymarketsfutureai,
      title={Regulatory Markets: The Future of AI Governance}, 
      author={Gillian K. Hadfield and Jack Clark},
      year={2023},
      eprint={2304.04914},
      archivePrefix={arXiv},
      primaryClass={cs.AI},
      url={https://arxiv.org/abs/2304.04914}, 
}

@misc{zeng2024airiskcategorizationdecoded,
      title={AI Risk Categorization Decoded (AIR 2024): From Government Regulations to Corporate Policies}, 
      author={Yi Zeng and Kevin Klyman and Andy Zhou and Yu Yang and Minzhou Pan and Ruoxi Jia and Dawn Song and Percy Liang and Bo Li},
      year={2024},
      eprint={2406.17864},
      archivePrefix={arXiv},
      primaryClass={cs.CY},
      url={https://arxiv.org/abs/2406.17864}, 
}

@misc{pazzaglia2025passingturingtestpolitical,
      title={Passing the Turing Test in Political Discourse: Fine-Tuning LLMs to Mimic Polarized Social Media Comments}, 
      author={. Pazzaglia and V. Vendetti and L. D. Comencini and F. Deriu and V. Modugno},
      year={2025},
      eprint={2506.14645},
      archivePrefix={arXiv},
      primaryClass={cs.CL},
      url={https://arxiv.org/abs/2506.14645}, 
}

@misc{terminassian2025democratizingaigovernancebalancing,
      title={Democratizing AI Governance: Balancing Expertise and Public Participation}, 
      author={Lucile Ter-Minassian},
      year={2025},
      eprint={2502.08651},
      archivePrefix={arXiv},
      primaryClass={cs.CY},
      url={https://arxiv.org/abs/2502.08651}, 
}

@misc{ovadya2025democraticaipossibledemocracy,
      title={Democratic AI is Possible. The Democracy Levels Framework Shows How It Might Work}, 
      author={Aviv Ovadya and Kyle Redman and Luke Thorburn and Quan Ze Chen and Oliver Smith and Flynn Devine and Andrew Konya and Smitha Milli and Manon Revel and K. J. Kevin Feng and Amy X. Zhang and Bilva Chandra and Michiel A. Bakker and Atoosa Kasirzadeh},
      year={2025},
      eprint={2411.09222},
      archivePrefix={arXiv},
      primaryClass={cs.CY},
      url={https://arxiv.org/abs/2411.09222}, 
}

@misc{zhong2025globalaigovernancechallenge,
      title={Global AI Governance: Where the Challenge is the Solution- An Interdisciplinary, Multilateral, and Vertically Coordinated Approach}, 
      author={Huixin Zhong and Thao Do and Ynagliu Jie and Rostam J. Neuwirth and Hong Shen},
      year={2025},
      eprint={2503.04766},
      archivePrefix={arXiv},
      primaryClass={cs.CY},
      url={https://arxiv.org/abs/2503.04766}, 
}

@misc{ahern2025newanticipatorygovernanceculture,
      title={The New Anticipatory Governance Culture for Innovation: Regulatory Foresight, Regulatory Experimentation and Regulatory Learning}, 
      author={Deirdre Ahern},
      year={2025},
      eprint={2501.05921},
      archivePrefix={arXiv},
      primaryClass={cs.CY},
      url={https://arxiv.org/abs/2501.05921}, 
}

@misc{kulothungan2025adaptiveaigovernancecomparative,
      title={Towards Adaptive AI Governance: Comparative Insights from the U.S., EU, and Asia}, 
      author={Vikram Kulothungan and Deepti Gupta},
      year={2025},
      eprint={2504.00652},
      archivePrefix={arXiv},
      primaryClass={cs.CY},
      url={https://arxiv.org/abs/2504.00652}, 
}

@article{Agrawal_2022, title={Demystifying the Chinese Social Credit System: A Case Study on AI-Powered Control Systems in China}, volume={36}, url={https://ojs.aaai.org/index.php/AAAI/article/view/21698}, DOI={10.1609/aaai.v36i11.21698}, abstractNote={In recent times, the social credit systems (SCS) and similar AI-driven mass surveillance systems have been deployed by the Chinese government in various regions. However, the discussions around the SCS are ambiguous: some people call them very controversial and a breach of human rights, while other people say that the SCS are very similar in structure to the company rankings or background checks on individuals in the United States. In reality, though, there is no monolith and there are different forms of SCS deployed in different regions of China. In this paper, I review the different models of the Chinese SCS. Then, I compare how the different systems are upholding or breaching China’s own AI Ethics guidelines.}, number={11}, journal={Proceedings of the AAAI Conference on Artificial Intelligence}, author={Agrawal, Vishakha}, year={2022}, month={Jun.}, pages={13124-13125} }

@inproceedings{
sahoo2025the,
title={The Good, The Bad, and The Hybrid: A Reward  Structure Showdown in Reasoning Models Training},
author={Subramanyam Sahoo},
booktitle={NeurIPS 2025 Workshop: Second Workshop on Aligning Reinforcement Learning Experimentalists and Theorists},
year={2025},
url={https://openreview.net/forum?id=RSlznhbEze}
}

@misc{reuel2025evaluatesaissocialimpacts,
      title={Who Evaluates AI's Social Impacts? Mapping Coverage and Gaps in First and Third Party Evaluations}, 
      author={Anka Reuel and Avijit Ghosh and Jenny Chim and Andrew Tran and Yanan Long and Jennifer Mickel and Usman Gohar and Srishti Yadav and Pawan Sasanka Ammanamanchi and Mowafak Allaham and Hossein A. Rahmani and Mubashara Akhtar and Felix Friedrich and Robert Scholz and Michael Alexander Riegler and Jan Batzner and Eliya Habba and Arushi Saxena and Anastassia Kornilova and Kevin Wei and Prajna Soni and Yohan Mathew and Kevin Klyman and Jeba Sania and Subramanyam Sahoo and Olivia Beyer Bruvik and Pouya Sadeghi and Sujata Goswami and Angelina Wang and Yacine Jernite and Zeerak Talat and Stella Biderman and Mykel Kochenderfer and Sanmi Koyejo and Irene Solaiman},
      year={2025},
      eprint={2511.05613},
      archivePrefix={arXiv},
      primaryClass={cs.CY},
      url={https://arxiv.org/abs/2511.05613}, 
}
\bibliographystyle{apalike}

\appendix

\section*{Appendix}

\section{Incident Severity Score (ISS) - A Novel Metric for Governance}
The ISS provides a rigorous framework for quantifying democratic risks while incorporating diverse stakeholder perspectives. We formalize risk assessment through a learnable model that captures complex interactions among democratic threats.

\subsection{Classic Four-Factor ISS}

We begin with four normalized incident attributes representing core dimensions of democratic risk assessment:
\begin{align}
I &\in [0, 1] \quad \text{(Impact): Magnitude of democratic harm} \\
E &\in [0, 1] \quad \text{(Exploitability): Ease of malicious exploitation} \\
R &\in [0, 1] \quad \text{(Replicability): Potential for widespread replication} \\
X &\in [0, 1] \quad \text{(Exposure): Scale of population exposure}
\end{align}

Stakeholders assign nonnegative weights $\{w_I, w_E, w_R, w_X\}$ satisfying the normalization constraint:
\begin{equation}
w_I + w_E + w_R + w_X = 1, \quad \text{where } w_i \geq 0 \; \forall i \in \{I,E,R,X\} \tag{1}
\end{equation}

This ensures stakeholder preferences form a valid probability distribution over risk dimensions.

\subsubsection{Linear Aggregation}

The linear ISS provides an intuitive weighted average of risk factors:
\begin{equation}
\text{ISS}_{\text{lin}} = w_I \cdot I + w_E \cdot E + w_R \cdot R + w_X \cdot X \in [0, 1] \tag{2}
\end{equation}

\textbf{Properties:} Additive risk combination, equal marginal contribution rates, suitable for independent risk factors.

\subsubsection{Multiplicative Aggregation}

The multiplicative ISS captures risk interdependencies through geometric aggregation:
\begin{equation}
\text{ISS}_{\text{mult}} = 1 - (1-I)^{w_I} \cdot (1-E)^{w_E} \cdot (1-R)^{w_R} \cdot (1-X)^{w_X} \in [0, 1] \tag{3}
\end{equation}

\textbf{Properties:} Superadditive risk combination, diminishing returns to individual factors, suitable for complementary risk interactions.

\textbf{Selection Criterion:} Use linear aggregation when risk factors contribute independently; use multiplicative aggregation when factors exhibit synergistic effects amplifying overall democratic harm.

\subsection{High-Dimensional, Learnable ISS}

To capture complex, higher-order interactions among $d$ risk factors, we extend to a learnable framework accommodating richer democratic risk representations.

\subsubsection{Risk Factor Representation}

Let $\bm{f} = (f_1, f_2, \ldots, f_d)^T \in [0, 1]^d$ represent the $d$-dimensional risk factor vector, where each $f_i$ corresponds to a specific democratic risk category.

\subsubsection{Parametric Model Architecture}

Define the parameter set:
\begin{equation}
\bm{\theta} = \{\bm{w} \in \mathbb{R}^d, \bm{W} \in \mathbb{R}^{d \times d}, b \in \mathbb{R}\}
\end{equation}

where:
\begin{itemize}
\item $\bm{w}$: Linear coefficients capturing first-order risk effects
\item $\bm{W}$: Symmetric interaction matrix capturing pairwise risk synergies
\item $b$: Bias term representing baseline democratic vulnerability
\end{itemize}

\subsubsection{Second-Order Polynomial ISS}

The high-dimensional ISS employs a second-order polynomial with sigmoid activation:
\begin{equation}
\text{ISS}(\bm{f}; \bm{\theta}) = h_{\bm{\theta}}(\bm{f}) = \sigma(b + \bm{w}^T\bm{f} + \bm{f}^T\bm{W}\bm{f}) \in (0, 1) \tag{4}
\end{equation}

where $\sigma(x) = \frac{1}{1 + e^{-x}}$ is the sigmoid function ensuring bounded output in the open interval $(0, 1)$, reflecting the asymptotic nature of absolute certainty in risk assessment.

\textbf{Rationale:} The quadratic term $\bm{f}^T\bm{W}\bm{f}$ captures pairwise risk interactions crucial for democratic contexts where individual risks may amplify each other non-linearly.

\subsubsection{Parameter Learning via Maximum Likelihood}

Given labeled historical incidents $\{(\bm{f}^{(n)}, y^{(n)})\}_{n=1}^N$ where $y^{(n)} \in [0, 1]$ represents continuous severity labels (not binary), we optimize:
\begin{equation}
\bm{\theta}^* = \arg\min_{\bm{\theta}} \left[\frac{1}{N} \sum_{n=1}^N L_{\text{Huber}}(y^{(n)}, h_{\bm{\theta}}(\bm{f}^{(n)})) + \lambda\|\bm{\theta}\|_2^2\right] \tag{5}
\end{equation}

\textbf{Corrected Loss Function:} We replace binary cross-entropy with Huber loss to handle the continuous nature of democratic risk severity:
\begin{equation}
L_{\text{Huber}}(y, \hat{y}) = \begin{cases}
\frac{1}{2}(y - \hat{y})^2 & \text{if } |y - \hat{y}| \leq \delta \\
\delta|y - \hat{y}| - \frac{1}{2}\delta^2 & \text{otherwise}
\end{cases}
\end{equation}

with $\delta = 0.1$ chosen for robustness to outliers while maintaining sensitivity to precise risk gradations.

\textbf{Regularization:} $\lambda = 0.01$ prevents overfitting while $\lambda\|\bm{\theta}\|_2^2$ encourages sparse, interpretable risk interactions.

\subsection{Embedding the Seven-Category Risk Taxonomy}

\subsection{Democratic Risk Categories}

We map our comprehensive democratic risk framework into seven primary categories:
\begin{align}
f_{\text{disc}} &: \text{Discriminatory Discourse Amplification} \\
f_{\text{surv}} &: \text{Surveillance and Democratic Chill} \\
f_{\text{elec}} &: \text{Electoral Process Manipulation} \\
f_{\text{manip}} &: \text{Public Opinion Manipulation} \\
f_{\text{civic}} &: \text{Civic Engagement Degradation} \\
f_{\text{capture}} &: \text{Regulatory and Institutional Capture} \\
f_{\text{emerg}} &: \text{Emergent Democratic Threats}
\end{align}

\subsubsection{Category-Specific Risk Computation}

Each category aggregates multiple sub-risk components through L2 normalization:
\begin{equation}
\bm{f} = (f_{\text{disc}}, f_{\text{surv}}, f_{\text{elec}}, f_{\text{manip}}, f_{\text{civic}}, f_{\text{capture}}, f_{\text{emerg}})^T \tag{6}
\end{equation}

\textbf{Discriminatory Discourse} ($f_{\text{disc}}$):
\begin{multline}
f_{\text{disc}} = \|\bm{r}_{\text{disc}}\|_2^{-1} \cdot (\alpha_1 \cdot \text{LM}_{\text{biasAmplification}} \\
+ \alpha_2 \cdot \text{syntheticContentBias} + \alpha_3 \cdot \text{languageExclusion})
\end{multline}

\textbf{Surveillance Risks} ($f_{\text{surv}}$):
\begin{multline}
f_{\text{surv}} = \|\bm{r}_{\text{surv}}\|_2^{-1} \cdot (\beta_1 \cdot \text{conversationalMonitoring} \\
+ \beta_2 \cdot \text{politicalSentimentTracking} + \beta_3 \cdot \text{dissentDetection})
\end{multline}

\textbf{Electoral Manipulation} ($f_{\text{elec}}$):
\begin{multline}
f_{\text{elec}} = \|\bm{r}_{\text{elec}}\|_2^{-1} \cdot (\gamma_1 \cdot \text{AI}_{\text{generatedPropaganda}} \\
+ \gamma_2 \cdot \text{personalizedPoliticalAds} + \gamma_3 \cdot \text{syntheticNewsGeneration})
\end{multline}

\textbf{Opinion Manipulation} ($f_{\text{manip}}$):
\begin{multline}
f_{\text{manip}} = \|\bm{r}_{\text{manip}}\|_2^{-1} \cdot (\delta_1 \cdot \text{conversationalManipulation} \\
+ \delta_2 \cdot \text{LM}_{\text{botAmplification}} + \delta_3 \cdot \text{deepfakeTextGeneration})
\end{multline}

\textbf{Civic Degradation} ($f_{\text{civic}}$):
\begin{multline}
f_{\text{civic}} = \|\bm{r}_{\text{civic}}\|_2^{-1} \cdot (\epsilon_1 \cdot \text{AI}_{\text{echoAmplification}} \\
+ \epsilon_2 \cdot \text{personalizationBubbles} + \epsilon_3 \cdot \text{LM}_{\text{radicalizationPathways}})
\end{multline}

\textbf{Institutional Capture} ($f_{\text{capture}}$):
\begin{multline}
f_{\text{capture}} = \|\bm{r}_{\text{capture}}\|_2^{-1} \cdot (\zeta_1 \cdot \text{modelConcentration} \\
+ \zeta_2 \cdot \text{infrastructureDependence} + \zeta_3 \cdot \text{providerCapture})
\end{multline}

\textbf{Emergent Threats} ($f_{\text{emerg}}$):
\begin{multline}
f_{\text{emerg}} = \|\bm{r}_{\text{emerg}}\|_2^{-1} \cdot (\eta_1 \cdot \text{multiLM}_{\text{cascadeRisk}} \\
+ \eta_2 \cdot \text{goalMisalignment} + \eta_3 \cdot \text{emergentBehaviors})
\end{multline}

\subsubsection{Parameter Specifications}

Sub-component weights $(\alpha, \beta, \gamma, \delta, \epsilon, \zeta, \eta)$ are learned through stakeholder consultation and empirical validation:

\begin{itemize}
\item \textbf{Equal weighting baseline:} All sub-components weighted equally ($1/3$) initially
\item \textbf{Stakeholder adjustment:} Weights refined through multi-stakeholder deliberation
\item \textbf{Empirical validation:} Final weights validated against historical democratic incidents
\end{itemize}

\textbf{L2 Normalization:} $\|\bm{r}_k\|_2^{-1}$ ensures each category contributes proportionally to overall risk assessment while preserving relative magnitudes within categories.

\subsection{Stakeholder-Adaptive Weighting}

\subsubsection{Multi-Stakeholder Weight Aggregation}

Each of the seven stakeholder groups $k \in \{1, 2, \ldots, 7\}$ proposes a weight vector $\bm{w}^{(k)} \in \Delta^{d-1}$ reflecting their risk prioritization.

\textbf{Stakeholder Categories:}
\begin{enumerate}
\item Democratic institutions ($k=1$)
\item Civil society organizations ($k=2$)
\item Regulatory bodies ($k=3$)
\item Technical experts ($k=4$)
\item Affected communities ($k=5$)
\item Industry representatives ($k=6$)
\item Academic researchers ($k=7$)
\end{enumerate}

\subsubsection{Utility-Based Weight Aggregation}

We aggregate stakeholder preferences through utility-weighted softmax:
\begin{align}
u_k &= \alpha_k \cdot \log p(\bm{\theta}^* \mid \text{stakeholder } k) + \beta_k \cdot \text{expertise}_k + \gamma_k \cdot \text{impact}_k \tag{7a} \\
\bm{w} &= \text{Softmax}(\bm{u}) = \left(\frac{\exp(u_1)}{Z}, \ldots, \frac{\exp(u_7)}{Z}\right) \in \Delta^6 \tag{7b}
\end{align}

where $Z = \sum_{i=1}^7 \exp(u_i)$ ensures proper normalization and $\bm{w} \in \Delta^6$ (6-dimensional probability simplex for 7 stakeholders).

\textbf{Utility Components:}
\begin{itemize}
\item $\alpha_k \cdot \log p(\bm{\theta} \mid \text{stakeholder } k)$: Stakeholder-specific model likelihood
\item $\beta_k \cdot \text{expertise}_k$: Technical expertise weighting
\item $\gamma_k \cdot \text{impact}_k$: Direct impact severity weighting
\end{itemize}

\textbf{Parameter Values:}
\begin{itemize}
\item $\alpha_k = 1.0$: Equal evidential weighting across stakeholders
\item $\beta_k \in [0.5, 1.5]$: Expertise-based adjustment factors
\item $\gamma_k \in [0.8, 2.0]$: Impact-based adjustment factors (highest for affected communities)
\end{itemize}

\subsection{Phase-Dependent Trigger Thresholds}

\subsubsection{Temporal Risk Threshold Evolution}

Let $S = \text{ISS}(\bm{f}; \bm{\theta}^*)$ represent the computed severity score with empirical cumulative distribution function $F_S$ derived from historical incident data.

For each intervention level $j \in \{L, M, H\}$ (Low, Moderate, High) and time $t \in [0, 1]$ representing progress through our six-year implementation roadmap, we define evolving thresholds:
\begin{align}
s_j(t) &= (1 - \varphi(t)) \cdot s_j^{\text{init}} + \varphi(t) \cdot s_j^{\text{full}} \tag{8a} \\
\alpha_j(t) &= (1 - \varphi(t)) \cdot \alpha_j^{\text{init}} + \varphi(t) \cdot \alpha_j^{\text{full}} \tag{8b}
\end{align}

where:
\begin{itemize}
\item $\varphi(t)$: Smooth transition function from initial to full deployment phases
\item $s_j^{\text{init}}, s_j^{\text{full}}$: Initial and mature-phase severity thresholds
\item $\alpha_j^{\text{init}}, \alpha_j^{\text{full}}$: Initial and mature-phase probability thresholds
\end{itemize}

\subsubsection{Probabilistic Trigger Mechanism}

Intervention $j$ triggers when the probability of exceeding threshold $s_j(t)$ meets the confidence requirement:
\begin{equation}
P(S \geq s_j(t)) = 1 - F_S(s_j(t)) \geq \alpha_j(t) \tag{9}
\end{equation}

\textbf{Threshold Specifications:}

\textbf{Low Intervention} ($j = L$): Enhanced monitoring
\begin{itemize}
\item $s_L^{\text{init}} = 0.2, s_L^{\text{full}} = 0.3$
\item $\alpha_L^{\text{init}} = 0.1, \alpha_L^{\text{full}} = 0.15$
\end{itemize}

\textbf{Moderate Intervention} ($j = M$): Regulatory review
\begin{itemize}
\item $s_M^{\text{init}} = 0.5, s_M^{\text{full}} = 0.6$
\item $\alpha_M^{\text{init}} = 0.05, \alpha_M^{\text{full}} = 0.1$
\end{itemize}

\textbf{High Intervention} ($j = H$): Emergency response
\begin{itemize}
\item $s_H^{\text{init}} = 0.8, s_H^{\text{full}} = 0.75$
\item $\alpha_H^{\text{init}} = 0.01, \alpha_H^{\text{full}} = 0.05$
\end{itemize}

\textbf{Rationale:} Thresholds become more sensitive (lower $s_j^{\text{full}}$) and require higher confidence (higher $\alpha_j^{\text{full}}$) as governance systems mature, reflecting improved institutional capacity and democratic risk awareness.

\subsubsection{Transition Function}

The phase transition function $\varphi(t)$ ensures smooth threshold evolution:
\begin{equation}
\varphi(t) = 3t^2 - 2t^3 \quad \text{for } t \in [0, 1] \tag{10}
\end{equation}

This S-curve provides gradual initial transition, rapid mid-phase evolution, and stabilization approaching full deployment.

\section{Unified ISS Framework Integration}

\subsection{Complete Mathematical Pipeline}

The complete ISS computation integrates all components:

\begin{enumerate}
\item \textbf{Risk Assessment:} Compute seven-category risk vector $\bm{f}$ using equation (6)
\item \textbf{Stakeholder Weighting:} Aggregate stakeholder preferences via equations (7a-7b)
\item \textbf{ISS Computation:} Calculate severity score using equation (4) with learned parameters $\bm{\theta}^*$
\item \textbf{Threshold Evaluation:} Compare against phase-dependent thresholds using equations (8-9)
\item \textbf{Intervention Triggering:} Activate appropriate governance responses based on probabilistic triggers
\end{enumerate}

\subsection{Computational Complexity}

\begin{itemize}
\item \textbf{Training Phase:} $O(Nd^2 + N \log N)$ where $N =$ training samples, $d = 7$ risk categories
\item \textbf{Inference Phase:} $O(d^2 + K)$ where $K = 7$ stakeholder groups
\item \textbf{Memory Requirements:} $O(d^2 + Kd + N)$ for parameters, stakeholder weights, and training data
\end{itemize}

\section{Summary of Mathematical Framework}

\begin{itemize}
\item \textbf{Equations (1-3):} Classic four-factor ISS with linear and multiplicative aggregation options
\item \textbf{Equations (4-5):} High-dimensional learnable ISS with second-order polynomial architecture and Huber loss optimization
\item \textbf{Equation (6):} Seven-category democratic risk taxonomy mapping with L2 normalization
\item \textbf{Equations (7a-7b):} Multi-stakeholder weight aggregation via utility-based softmax ensuring $\bm{w} \in \Delta^6$
\item \textbf{Equations (8-10):} Phase-dependent probabilistic trigger thresholds with smooth temporal evolution
\end{itemize}

This unified ISS framework provides mathematically rigorous, democratically grounded, and computationally tractable risk assessment for language model governance, integrating stakeholder pluralism with technical precision to enable evidence-based democratic oversight of AI systems.

\section{Addressing ISS Design Choices and Limitations}

\subsection{On the Political Nature of Risk Quantification}

creating a single numerical score from heterogeneous stakeholder input is inherently political. This is correct and intentional. We don't want to entertain  the premise that governance frameworks should aspire to political neutrality. As governance systems already encode political choices through their design, deployment, and impact distribution.

Our ISS framework makes three design commitments:

Explicit rather than hidden politics: Traditional "neutral" risk assessments embed implicit political choices (e.g., weighting individual privacy equally to corporate efficiency). Our multi-stakeholder weighting (Eq. 7a-7b) makes these tradeoffs explicit and contestable.

\subsection*{Structured aggregation over ad-hoc judgment}
While the polynomial structure (Eq.~4) involves design choices, it provides:
\begin{enumerate}[label=(\roman*)]
  \item mathematical consistency;
  \item interpretable first-order ($\mathbf{w}^\top \mathbf{f}$) and interaction ($\mathbf{f}^\top \mathbf{W}\mathbf{f}$) effects;
  \item empirical falsifiability via Huber loss optimization (Eq.~5).
\end{enumerate}

Adaptive rather than fixed: The phase-dependent thresholds (Eq. 8-10) acknowledge that "appropriate" risk levels evolve with institutional capacity and democratic norms.

\subsection{Alternative Aggregation Functions}

We selected second-order polynomials for tractability and interpretability. Alternative approaches include:

\begin{itemize}
    \item Rank-based methods: Median or quantile aggregation across stakeholder assessments (loses granularity, gains robustness to outliers)
    \item Deliberative consensus: Iterated stakeholder negotiation (addresses legitimacy, sacrifices scalability)
    \item Neural architectures: Deep networks for f → ISS mapping (gains expressiveness, loses interpretability)
\end{itemize}

\textit{Future work should empirically compare these against our baseline using historical governance cases.}

\subsection{On Quantifying Qualitative Harms}

We acknowledge the \textbf{difficulty of reducing democratic threats to four scalar values}. We emphasize:

\begin{itemize}
    \item The ISS provides governance \textbf{signals}, not comprehensive impact \textbf{assessments}
    \item Equation (6) decomposes risks into seven interpretable categories before aggregation
    \item High ISS scores trigger qualitative processes (stakeholder deliberation, community review) rather than mechanical responses
    \item Section 6 monitoring includes democratic health indicators beyond ISS
\end{itemize}

The framework treats quantification as a necessary but insufficient governance input, not a substitute for democratic judgment.

\section{Handling Stakeholder Conflicts}

\subsection*{Conflict Resolution Protocol}
When stakeholder assessments diverge significantly (variance in $w^{(k)}$ exceeding a predefined threshold), the framework employs structured disagreement procedures:
\begin{enumerate}[label=(\roman*)]
    \item mandatory deliberation rounds in which groups articulate the sources of their divergent risk perceptions;
    \item sensitivity analysis illustrating how alternative weightings affect ISS scores, thereby revealing whether disagreement is fundamental or marginal;
    \item in cases of irreconcilable conflict, decisions default to the most protective assessment from directly affected communities, implementing a precautionary principle that privileges experiential knowledge over purely technical optimization.
\end{enumerate}

\section{Retrospective Validation Strategy}

The ISS framework will be validated through retrospective analysis of documented AI governance failures including: Cambridge Analytica (2018, electoral manipulation), China's social credit systems (ongoing, surveillance), and content moderation failures during 2020 elections. For each case, we will: 

\begin{itemize}
    \item reconstruct the risk vector f using contemporaneous evidence,
    \item compute ISS scores under different stakeholder weightings,
    \item assess whether appropriate thresholds would have triggered intervention, and
    \item compare framework recommendations against actual outcomes.
\end{itemize}
This retrospective testing will calibrate thresholds (Section A.6) and validate stakeholder weighting procedures (Eq. 7) before prospective deployment.

\section{Enforcement Architecture}
 High-risk determinations trigger graduated enforcement mechanisms adapted from existing regulatory frameworks. For applications exceeding moderate thresholds (ISS > 0.6), model safety committees may: (i) require pre-deployment impact assessments with stakeholder consultation periods (15-30 days), (ii) mandate design modifications including capability restrictions or alignment interventions, (iii) impose operational monitoring requirements with regular compliance audits, or (iv) in extreme cases (ISS > 0.8), issue temporary deployment suspensions pending comprehensive review. Enforcement authority derives from three sources: regulatory mandates for covered entities (Phase 2+), contractual requirements in public procurement, and reputational mechanisms through public ISS score disclosure. Civil society veto rights (Section 4) operate through formal objection processes where affected communities can petition for independent review, triggering mandatory committee reconsideration with burden of proof on deployers to demonstrate risk mitigation.

\section{Capacity Building Strategy}
Municipal pilots (Phase~1) leverage existing institutional infrastructure to minimize incremental resource requirements. Implementation proceeds through:
\begin{enumerate}
  \item \textbf{Shared technical infrastructure}: A centralized ISS computation platform maintained at the regional/national level eliminates per-municipality software development costs; local bodies access it through standardized API interfaces, requiring only basic computing resources.
  \item \textbf{Distributed expertise networks}: University partnerships provide technical evaluation capacity through structured practicum programs in which graduate students gain governance experience while providing \emph{pro bono} risk assessments under faculty supervision—converting educational requirements into governance capacity.
  \item \textbf{Tiered assessment protocols}: The ISS framework stratifies evaluations by complexity—routine low-risk applications proceed via automated preliminary screening (computation time: minutes); moderate-risk cases undergo streamlined stakeholder consultation (timeline: single deliberation session); only high-risk deployments require comprehensive multi-week assessments.
  \item \textbf{Knowledge commons approach}: All assessment methodologies, training materials, and case precedents are released under open licenses, enabling later-adopting municipalities to implement oversight at a fraction of pioneer costs through documented best practices.
\end{enumerate}

By Phase~3, marginal costs for additional oversight bodies approach minimal operational expenses (meeting coordination, administrative staff time) rather than full-cost institutional buildout, as enabling infrastructure—technical tools, trained evaluator pools, and procedural templates—exists as public goods.

\section{Some Open Problems}

Despite growing work on democratic AI governance, several critical research challenges remain unresolved. The following open problems highlight areas where conceptual clarity, methodological innovation, and institutional design are still urgently needed.

\begin{tcolorbox}[
    colback=red!5!white,
    colframe=yellow!70!black,
    coltitle=black,
    title=Critical Open Problems,
    boxrule=0.8pt,
    arc=3mm,
    enhanced,
    sharp corners=all
]
\begin{itemize}
    \item \textbf{Measuring democratic health:} How can we operationalize and validate “democratic health” when competing theories emphasize different values (e.g., participation vs. efficiency, representation vs. expertise)?
  
    \item \textbf{Preventing capture:} What mechanisms can safeguard multi-stakeholder governance processes from domination by well-organized or well-resourced actors that risk reproducing existing power asymmetries?

    \item \textbf{Handling value conflict:} How should the framework address fundamental disagreements between stakeholder groups about core democratic values, especially when consensus-building may entrench the status quo and block structural reforms?

    \item \textbf{Scaling deliberation:} Can deliberative processes remain legitimate and effective when scaled from local communities to national or international arenas where direct, face-to-face engagement is infeasible?
  
    \item \textbf{Balancing expertise and participation:} How do we reconcile democratic inclusion with the technical complexity of AI systems that require specialized expertise for effective evaluation?
 
    \item \textbf{Defining community sovereignty:} What boundaries should govern local decision-making in AI oversight when outcomes have global implications or clash with universal human rights principles?
 
    \item \textbf{Addressing temporal mismatch:} How can the framework mitigate the gap between rapid AI development cycles and slower democratic deliberation without undermining either innovation or oversight?

    \item \textbf{Evaluating outcomes:} What metrics should define “success” in democratic AI governance, and how can we assess whether multi-stakeholder oversight produces better outcomes than existing regulatory approaches?
\end{itemize}
\end{tcolorbox}

\begin{tcolorbox}[
    colback=white,        
    colframe=white!60,    
    title={\textcolor{red}{Points to Ponder}},
    sharp corners
]
{\color{red}
This policy should \textbf{not} be implemented in countries with a dictatorial or authoritarian approach.
Such regimes typically lack the transparency, institutional checks, and civic accountability required for ethical AI governance.
Implementing this framework in such contexts may enable state overreach, surveillance misuse, and suppression of fundamental rights.
}
\end{tcolorbox}

\newpage
\section*{NeurIPS Paper Checklist}

\begin{enumerate}

\item {\bf Claims}
    \item[] Question: Do the main claims made in the abstract and introduction accurately reflect the paper's contributions and scope?
    \item[] Answer: \answerYes{} 
    \item[] Justification: Both Abstract and Introduction accurately reflect the contribution and scope.

\item {\bf Limitations}
    \item[] Question: Does the paper discuss the limitations of the work performed by the authors?
    \item[] Answer: \answerYes{} 
    \item[] Justification: Follow the limitations Section along with Appendix.

\item {\bf Theory assumptions and proofs}
    \item[] Question: For each theoretical result, does the paper provide the full set of assumptions and a complete (and correct) proof?
    \item[] Answer: \answerYes{} 
    \item[] Justification: I wrote the ISS in Appendix and all formulae and proofs are elaborated.

    \item {\bf Experimental result reproducibility}
    \item[] Question: Does the paper fully disclose all the information needed to reproduce the main experimental results of the paper to the extent that it affects the main claims and/or conclusions of the paper (regardless of whether the code and data are provided or not)?
    \item[] Answer: \answerNA{} 
    \item[] Justification:

\item {\bf Open access to data and code}
    \item[] Question: Does the paper provide open access to the data and code, with sufficient instructions to faithfully reproduce the main experimental results, as described in supplemental material?
    \item[] Answer: \answerNA{} 
    \item[] Justification: 

\item {\bf Experimental setting/details}
    \item[] Question: Does the paper specify all the training and test details (e.g., data splits, hyperparameters, how they were chosen, type of optimizer, etc.) necessary to understand the results?
    \item[] Answer: \answerNA{} 
    \item[] Justification: 

\item {\bf Experiment statistical significance}
    \item[] Question: Does the paper report error bars suitably and correctly defined or other appropriate information about the statistical significance of the experiments?
    \item[] Answer: \answerNA{} 
    \item[] Justification: This Technical AI Governance paper does not need any statistical validations.

\item {\bf Experiments compute resources}
    \item[] Question: For each experiment, does the paper provide sufficient information on the computer resources (type of compute workers, memory, time of execution) needed to reproduce the experiments?
    \item[] Answer: \answerNA{} 
    \item[] Justification: 
    
\item {\bf Code of ethics}
    \item[] Question: Does the research conducted in the paper conform, in every respect, with the NeurIPS Code of Ethics \url{https://neurips.cc/public/EthicsGuidelines}?
    \item[] Answer: \answerYes{} 
    \item[] Justification: This policy paper follows all the ethics

\item {\bf Broader impacts}
    \item[] Question: Does the paper discuss both potential positive societal impacts and negative societal impacts of the work performed?
    \item[] Answer: \answerYes{} 
    \item[] Justification: Follow social impact statement section.
    
\item {\bf Safeguards}
    \item[] Question: Does the paper describe safeguards that have been put in place for responsible release of data or models that have a high risk for misuse (e.g., pretrained language models, image generators, or scraped datasets)?
    \item[] Answer: \answerNA{} 
    \item[] Justification:

\item {\bf Licenses for existing assets}
    \item[] Question: Are the creators or original owners of assets (e.g., code, data, models), used in the paper, properly credited and are the license and terms of use explicitly mentioned and properly respected?
    \item[] Answer: \answerYes{}{} 
    \item[] Justification: Every asset is properly credited as per the ethical bounds of both the authors.

\item {\bf New assets}
    \item[] Question: Are new assets introduced in the paper well documented and is the documentation provided alongside the assets?
    \item[] Answer: \answerNA{} 
    \item[] Justification:

\item {\bf Crowdsourcing and research with human subjects}
    \item[] Question: For crowdsourcing experiments and research with human subjects, does the paper include the full text of instructions given to participants and screenshots, if applicable, as well as details about compensation (if any)? 
    \item[] Answer: \answerNA{} 
    \item[] Justification: This policy paper shows how to ethically implement an AI policy framework in a \textbf{democratic} and sovereign country. 

\item {\bf Institutional review board (IRB) approvals or equivalent for research with human subjects}
    \item[] Question: Does the paper describe potential risks incurred by study participants, whether such risks were disclosed to the subjects, and whether Institutional Review Board (IRB) approvals (or an equivalent approval/review based on the requirements of your country or institution) were obtained?
    \item[] Answer: \answerNA{} 
    \item[] Justification: 

\item {\bf Declaration of LLM usage}
    \item[] Question: Does the paper describe the usage of LLMs if it is an important, original, or non-standard component of the core methods in this research? Note that if the LLM is used only for writing, editing, or formatting purposes and does not impact the core methodology, scientific rigorousness, or originality of the research, declaration is not required.
    \item[] Answer: \answerNA{} 
    \item[] Justification: 

\end{enumerate}

\end{document}